\documentclass[authoryear]{elsarticle} 

\usepackage{lineno,hyperref}
\usepackage{amsmath}
\usepackage{setspace}
\usepackage[margin=0.75in]{geometry}
\usepackage{graphicx}
\usepackage{epstopdf}
\usepackage{amssymb}
\usepackage[dvipsnames]{xcolor}
\modulolinenumbers[1]
\usepackage{natbib}

\journal{}

\begin{document}
\begin{frontmatter}
\title{Chua Circuit based on the Exponential Characteristics of Semiconductor Devices}
\author{Ronilson Rocha (Orcid id 0000-0002-9079-8196)}
\address{(Corresponding Author)\\
           Federal University of Ouro Preto\\
           UFOP/EM/DEMEC\\
           Campus Morro do Cruzeiro, 35400-000, Ouro Preto, MG, Brazil\\
           email: rocha@ufop.edu.br}
\author{Rene Orlando Medrano-T (Orcid id 0000-0003-0866-2466)}
\address{Federal University of S\~ao Paulo\\
           UNIFESP/DF\\
           Campus Diadema, 09972-270, Diadema, SP, Brasil \\
	   S\~ao Paulo State University\\
           UNESP/IGCE/DF \\
           Campus Rio Claro, 13506-900, Rio Claro, SP, Brazil\\
	  email: rene.medrano@unifesp.br}

\begin{abstract}
 {The use of non-ideal features of semiconductor devices is an interesting option for implementations of nonlinear electronic systems.
This paper analyzes the Chua circuit with nonlinearity based on the exponential hyperbolic characteristics of semiconductor devices. 
The stability analysis using describing functions predicts the dynamics of this nonlinear system, which is corroborated by numerical investigations and experimental results. 
The dynamic behaviors and bifurcations of this nonlinear system are mapped in parameter space in order to create a base for studies, analyses, and designs.
The dynamic behavior of the experimental high speed implementation of this version of Chua circuit differs from the expected dynamics for a conventional Chua circuit due to effects of unmodelled non-idealities of the real semiconductor devices,displaying that new and different dynamics for the Chua circuit can be obtained exploring different nonlinearities.}
\end{abstract}

\begin{keyword}
Nonlinear System, Chua Circuit, Exponential Hyperbolic Function, Semiconductor Devices, Stability Analysis, Describing Functions.
\end{keyword}

\end{frontmatter}

\section{Introduction}
\begin{figure}
\centering
\small
\includegraphics[angle=0, height=3cm]{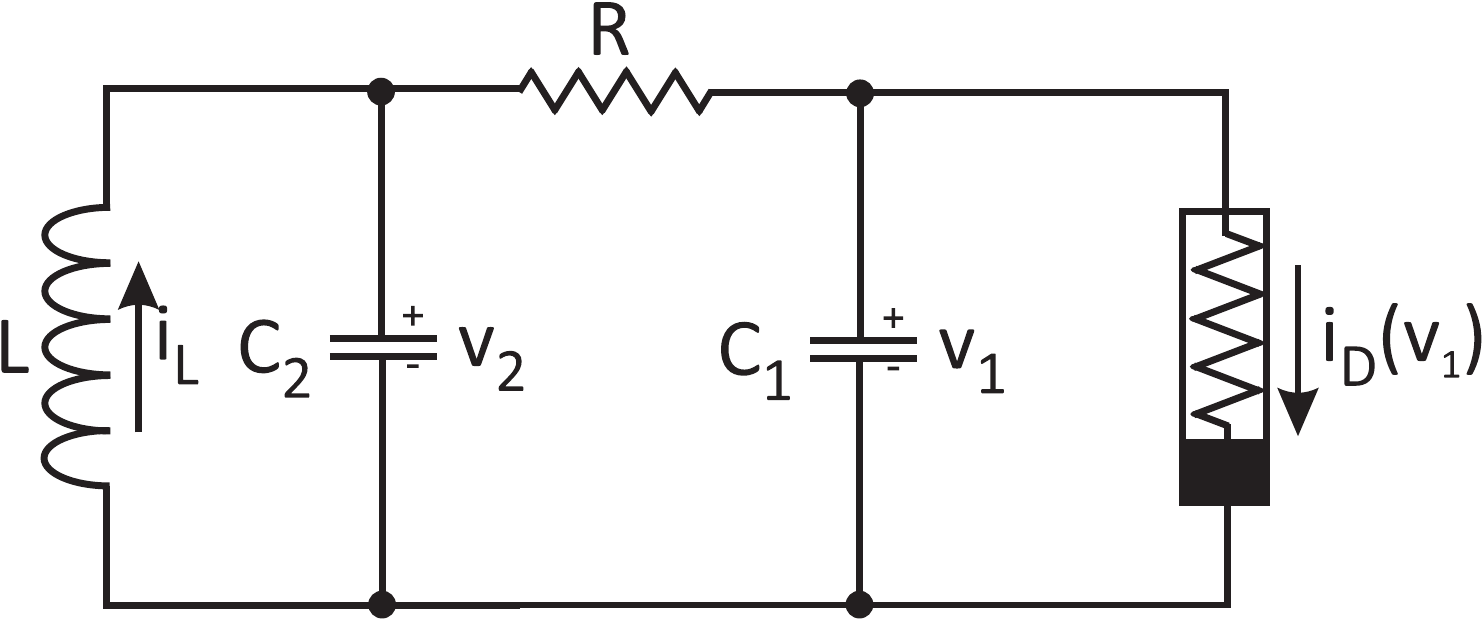}
\caption{The standard Chua circuit.}
\label{Fig:Chua_circuit}
\end{figure}

The Chua circuit is an important nonlinear oscillator whose standard form is presented in Fig. \ref{Fig:Chua_circuit}, where a nonlinear voltage controlled current source known as Chua diode supplies power to a passive low pass filter composed by the capacitor $C_1$ connected to the resonant circuit $LC_2$ through the linear resistor $R$.
The Chua circuit can eventually have additive branches with electro-electronic devices for capture and shunt energy in order to induce different operation modes \citep{wang-2020}.
Considering a breakpoint voltage $B$ as amplitude scale factor and the time constant $\tau=RC_2$ as time scale factor, the dynamics of original Chua circuit can be conveniently modelled using a reduced number of dimensionless parameters groups as

\begin{align}
\dot{x}&=\alpha\left[-x+y-u(x)\right], \nonumber \\
\dot{y}&=x-y+z, \label{eq-estados}\\
\dot{z}&=-\beta y \nonumber, 
\end{align}
where

\begin{equation}
x=\frac{v_1}{B}, \quad y=\frac{v_2}{B}, \quad z=\frac{Ri_L}{B},\quad \alpha=\frac{C_2}{C_1}, \quad \beta=\frac{R^2C_2}{L},\nonumber
\end{equation}
$v_1$ is the voltage in the capacitor $C_1$, $v_2$ is the voltage in the capacitor $C_2$, $i_L$ is the current in the inductor $L$, and $u(x)$ is the dimensionless output current $i_D(v_1)$ of the Chua diode given by

\begin{equation}
u(x)=\frac{R}{B}i_{D}(Bx).
\end{equation}
The nonlinear current-voltage characteristic $i_D(v_1)$ of the Chua diode is usually a three-segmented piecewise linear function \citep{chua-1986,brown-1993,tsuneda-2005,rocha-2009}.
Other nonlinearities have been proposed for Chua diode, such as cubic polynomial functions and  ``cubic-like'' approximations  \citep{zhong-1994,eltawil-1999,donoghue-2005,tsuneda-2005,rocha-2020}, sigmoid and signum functions \citep{brown-1993}, odd square law $ax+bx|x|$ \citep{tang-1998}, trigonometric functions \citep{tang-2001}, memristive current-voltage characteristics \citep{rocha-2017}, etc.
In despite of its simplicity, the Chua circuit generates a great diversity of nonlinear phenomena such as fixed and equilibrium points, periodic and stranger attractors, Andronov-Hopf, saddle-node (tangent), flip (period-doubling), cusp, homoclinic, heteroclinic, and other kinds of bifurcations, multistability and hidden oscillations, antiperiodic oscillations, period-adding in sets of periodicity, metamorphoses of basins of attraction, etc \citep{madan-1993,medrano-2005,algaba-2012,leonov-2013,medrano-2014,singla-2015,menacer-2016,bao-2016b,bao-2018,singla-2018,liu-2020,wang-2021}.
The most of these nonlinear phenomena occur in the parameter range $\alpha<\beta<\gamma^2$ \citep{rocha-2020,rocha-2015,rocha-2016,rocha-2017}, where

\begin{equation}
\gamma=\frac{1+\alpha}{2}.
\end{equation}

 {Several works have proposed the intentional use of non-idealities of electronic devices as nonlinearities for implementation of nonlinear circuits with complex dynamics \citep{ozoguz-2002,tamaseviciute_2008,sprott-2011,fouda_2015,buscarino-2016,pham-2016,njitacke-2017,volos-2017,bao-2018,liu-2018,fozin-2019,signing-2019}. 
Considering the electro-electronic nature of the Chua circuit, an interesting proposal for implementation of its nonlinear element is use the intrinsic exponential hyperbolic current-voltage characteristics of semiconductor devices such as rectifier diodes, Zener diodes, Schottky diodes, tunnel diodes, photodiodes, light-emitting diodes, bipolar junction transistors, junction field effect transistors, etc. 
This non-ideality of semiconductor devices are rarely used in theoretical nonlinear systems due to possible divergences in numerical computations \citep{sprott-2000,hanias-2011,hanias-2011-a,pham-2016,liu-2018}. 
Semiconductor devices also introduce other nonlinear features in electro-electronic implementations, such as hysteresis, time delays, nonlinear leakage resistance, nonlinear parasitic capacitance, nonlinear lead self-inductance, etc  \citep{hanias-2011,petrzela-2018}.
These electronic components can also present sensitivity to energy fields, which allows interactions with external energy fields without the necessity of additional branches in the Chua circuit.}

\begin{figure}
\centering
\small
\includegraphics[height=3cm]{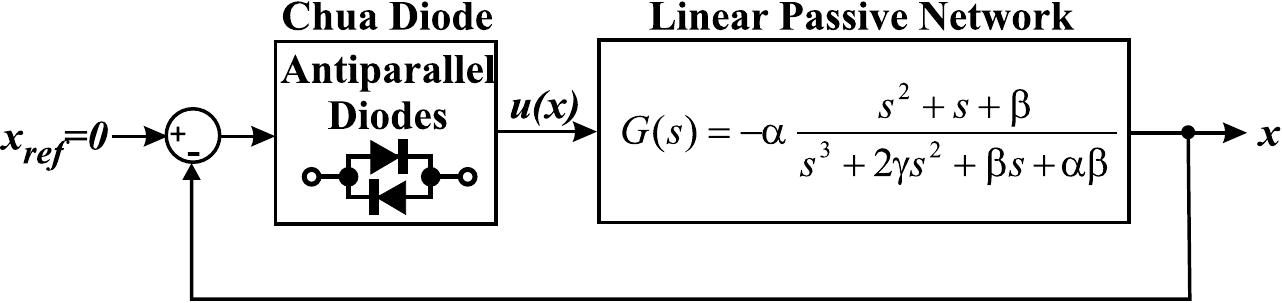}
\caption{The Chua circuit as a nonlinear feedback system.}
\label{Fig:Chua_feedback}
\end{figure}

The dynamic behavior of the Chua circuit depends on the parameters of the both passive low pass filter and nonlinear element, which can be mapped in parameter spaces using analytical-numerical approaches.
This mapping in parameter spaces is usually performed from exhaustive numerical investigations based on simulations and computation of the Lyapunov exponents \citep{komuro-1991,genesio-92-a,albuquerque-2008,stegemann-2010,albuquerque-2012,hoff-2014,medrano-2014}.
 {Since the Chua circuit is a nonlinear feedback system composed by a smooth nonlinearity associated to a linear low-pass filter that sufficiently attenuates higher-order harmonics as shown in Fig. \ref{Fig:Chua_feedback}, its dynamics can easily be predicted and mapped without complex computations using an analytical method known as describing functions.
This method allows understand and solve a large variety of problems related to analysis and design of nonlinear systems, providing reasonably accurate predictions for equilibrium points, nonlinear oscillations, periodic orbits, limit cycles, chaotic behavior, multistability and hidden attractors, subharmonics, jump resonance, and unstable behavior \citep{ogata-1970,genesio-91,slotine-1991,genesio-92,savaci-06,bragin-2011,medrano-2014,rocha-2015,rocha-2016,rocha-2017}. 
It is particularly useful to estimate the resulting effects of modifications in nonlinear systems, allowing find possible solutions in order to improve response and performance of a nonlinear plant. 
The approach using describing functions has a wide application in control theory for modelling and synthesis of linear and nonlinear compensators for control and stabilization of dynamic systems  \citep{ogata-1970,slotine-1991,genesio-93}.}

 {This paper analytically characterizes the dynamics of the Chua circuit with nonlinearity based on the inherent exponential hyperbolic characteristics of real semiconductor devices.
The dynamical behavior of this system is predicted from a stability analysis of equilibrium points and limit cycles using the method of the describing functions.
The dynamic behaviors and bifurcations of this nonlinear system are mapped in parameter space to create a base for studies, analyses, and designs. 
Numerical investigations and experimental results obtained in the low speed implementation of the Chua circuit corroborate the theoretical mapping. 
The effects of the unmodelled non-idealities of real semiconductor are experimentally observed in the fast speed implementation of this version of the Chua circuit, whose dynamic behavior differs from the expected dynamics of a conventional Chua circuit. 
This fact suggests that the use of unmodelled non-idealities of semiconductor devices could allow obtain new and different dynamics for the Chua circuit, which can consist in an attractive field for future researches.
The dynamical behavior of this system is predicted and mapped in parameter space from a stability analysis using the method of the describing functions in section \ref{sec:stability}, whose results are corroborated by numerical investigations in section \ref{sec:num}.
The description of electronic circuits for implementation of Chua circuit with nonlinearity based on the current-voltage characteristic of semiconductor devices and experimental results are presented in section \ref{sec:electronics}.
The conclusions are presented in section \ref{sec:conclusion}.}

\section{Stability Analysis} \label{sec:stability}
\begin{figure}
\centering
\small
\includegraphics[angle=0,height=4cm]{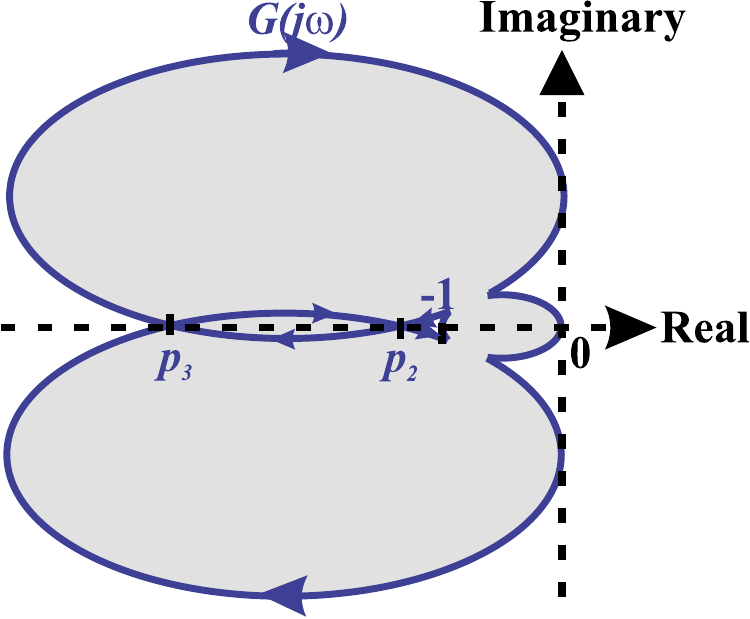}
\caption{The Nyquist diagram $G(j\omega)$ for the linear passive filter of the Chua circuit when $\alpha<\beta<\gamma^2$. Since $G(s)$ has no right half-plane poles for $\alpha>0$ and $\beta>0$, the Nyquist diagram $G(j\omega)$ establishes stable (white) and unstable (gray) regions for nonlinear feedback in the complex plane. The tendency of the signal amplitude is decrease in stable regions and increase in unstable regions.}
\label{Fig:nyquist}
\end{figure}

The transfer function $G(s)$ from the output $x$ to the input $u(x)$ for the linear passive filter of the Chua circuit is

\begin{equation}
G(s)=\frac{X(s)}{U(s)}=-\alpha \frac{s^2+s+\beta}{s^3+2\gamma s^2+\beta s+\alpha\beta}, \label{eq-tf}
\end{equation}
which presents a low-pass characteristic due to the excess of poles in relation to zeros.
The frequency response to sinusoidal signals of this linear passive filter for $\alpha<\beta<\gamma^2$ can be graphically represented by Nyquist diagram presented in Fig. \ref{Fig:nyquist}, which is the polar plotting of $G(j\omega)$ when the frequency $\omega$ is swept from $-\infty$ to $+\infty$.

The Shockley equation for $l$ parallel and $m$ series connected p-n semiconductor junctions is

\begin{equation}
i_d(v_d)=li_s\left(e^{\frac{v_d}{m\eta v_T}}-1\right),
\end{equation}
where $i_d$ is the direct diode current, $v_d$ is the diode voltage, $i_s$ is the reverse diode current, $v_T$ is the thermal voltage ($v_T \approx 26$ mV at room temperature), and $\eta$ is the emission coefficient that depends on the semiconductor material.
Thus, dimensionless characteristic $u(x)$ of a Chua diode composed by antiparallel association of semiconductor diodes with a parallel conductance $g_p$ is

\begin{equation}
u(x)= -\left[g_0x+I_0\sinh\left(x\right)\right], \label{eq:nonlinearity}
\end{equation}
where $B=m\eta v_T$, $g_0=Rg_p$ and $I_0=2Rli_s/B$.
Since the linear passive low-pass filter $G(s)$ attenuates higher frequency signals, the fundamental harmonic can be considered the only representative component in the output signal $u(x)$ such that this nonlinearity can be approximated by its describing function $N(X)$ given by

\begin{align}
N(X)&=-g_0-I_0\left\{1+\sum_{j=1}^\infty\left[\prod_{i=1}^j\left(\frac{2i+1}{2i+2}\right)\frac{X^{2j}}{(2j+1)!}\right]\right\},\label{eq:describing}
\end{align} 
where $X$ is the amplitude of the variable $x$.

\begin{figure}
\centering
\small
\includegraphics[angle=0,height=4cm]{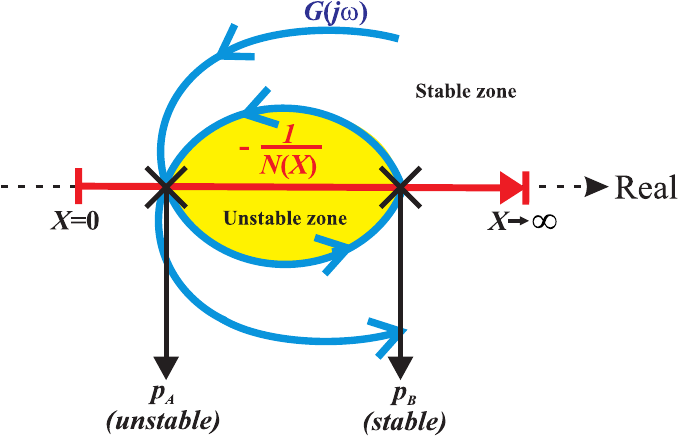} 
\caption{Stability of a limit cycle. The red line corresponds to the geometric locus $-1/N(X)$ while the blue curve represents a detail of the Nyquist diagram $G(j\omega)$, which encircles the unstable zone in yellow. The tendency is the amplitude $X$ decreases in the stable regions while it increases in unstable regions. Since the amplitude X diverges from $p_A$, this interception point corresponds to a unstable limit cycle. Considering the amplitude X converges to $p_B$, this interception point corresponds to a stable limit cycle.}
\label{limit-cycle}
\end{figure}

 {The stability analysis of nonlinear systems using describing functions is an extension of the famous Nyquist stability criterion, which establishes that the origin of the state-space in a nonlinear feedback system is stable only if the difference between conterclock and conterclockwise encirclements of the Nyquist diagram $G(j\omega)$ around the geometrical locus $-1/N(X)$ is equal to the quantity of poles with positive real part of $G(s)$.
Otherwise, the origin is unstable. 
Although this method explicitly verifies the stability of origin, it allows detect the existence and analyze the stability of limit cycles and equilibrium points out of the origin. 
A nonlinear feedback system has limit cycles if the geometric locus $-1/N(X)$ intercepts the Nyquist diagram $G(j\omega)$.
Equilibrium points out of the origin can be considered as limit cycles with null frequency. 
The stability of limit cycles and equilibrium points is analyzed according to Fig.  \ref{limit-cycle}.}
Superposing $-1/N(X)$ to $G(j\omega)$ in order to analyze the stability of the Chua circuit, the frequencies $\omega_i$ and the respective points $p_i$ where the geometric locus can intercept the Nyquist diagram are

\begin{align}
\omega_0=\pm\infty & \rightarrow p_0=0,\\
\omega_1=0 & \rightarrow p_1=-1,\\
\omega_2= \sqrt{\beta-\gamma-\sqrt{\gamma^2-\beta}} & \rightarrow p_2=-\frac{\alpha}{\gamma+\sqrt{\gamma^2-\beta}},\\
\omega_3= \sqrt{\beta-\gamma+\sqrt{\gamma^2-\beta}} & \rightarrow p_3=-\frac{\alpha}{\gamma-\sqrt{\gamma^2-\beta}}.
\end{align}

\begin{figure}
\centering
\small
\begin{tabular}{cc}
\includegraphics[height=3cm]{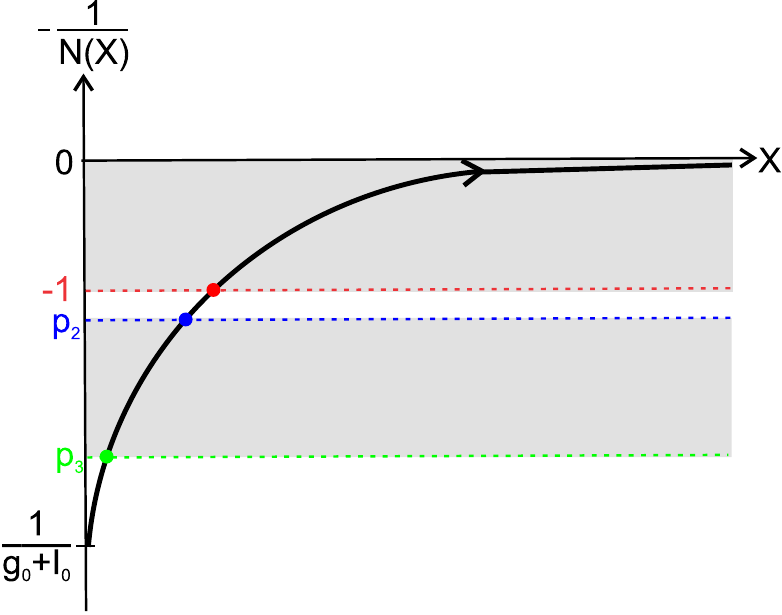} & \includegraphics[height=3cm]{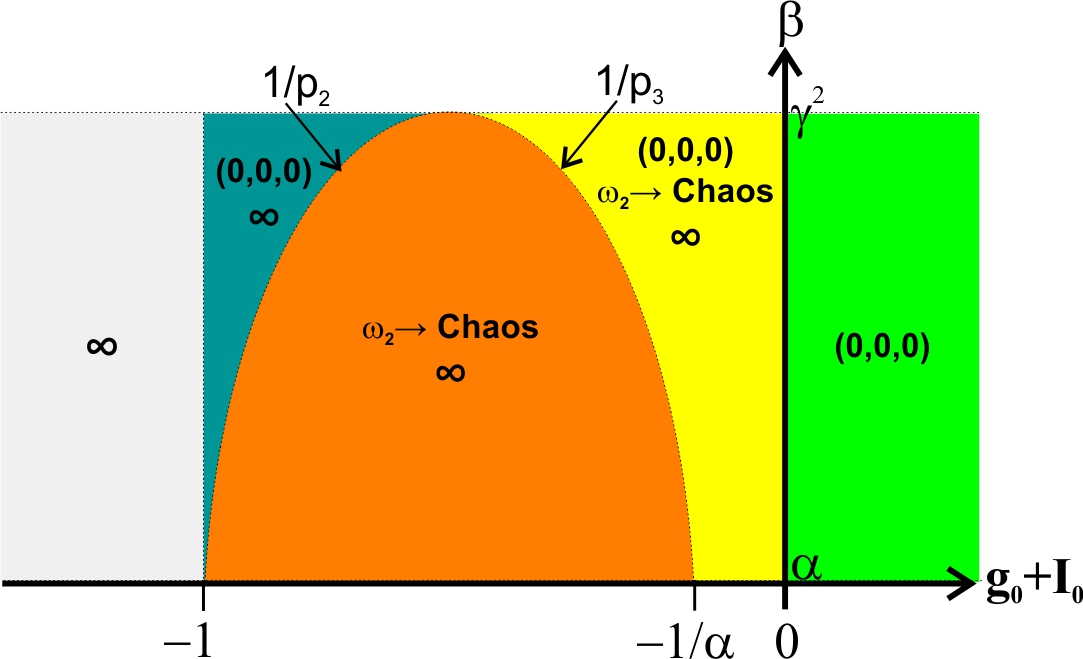}\\
(a) & (b)  
\end{tabular}
\caption{Stability analysis for $\alpha<\beta<\gamma^2$ and $(g_0+I_0) I_0>0$ using describing functions. (a) The geometric locus $-1/N(X)$ superposed to the Nyquist diagram $G(j\omega)$ and the three possible interception points: $\pm p_1$ (red circle), $p_2$ (blue circle), and $p_3$ (green circle). 
(b) Sketch of the theoretical parameter space mapping $(g_0+I_0) \times \beta$ based on the stability analysis, where $(0,0,0)=$ operation at origin $(0,0,0)$, $\omega_2\rightarrow Chaos=$ operation in a stable limit cycle with frequency $\omega_2$ that can evolve to chaos, and $\infty=$ unstable circuit.}
\label{Fig:stability}
\end{figure}

The describing function $N(x)$ has no real roots for $(g_0+I_0) I_0>0$ and the geometric locus $-1/N(X)$ is a continuous real function that starts at $-1/(g_0+I_0)$ for $X=0$ and ends at $0$ for $X=\infty$. 
The stability analysis for $\alpha<\beta<\gamma^2$ and $(g_0+I_0)I_0>0$ is presented in Fig. \ref{Fig:stability}.(a) and summarized in the sketch of the parameter space mapping presented in Fig. \ref{Fig:stability}.(b).
The operation of the Chua circuit is unstable for $-\infty<(g_0+I_0)<-1$ since $G(j\omega)$ completely involves $-1/N(x)$.
The geometric locus $-1/N(X)$ intercepts $G(j\omega)$ in the unstable equilibrium points $\pm p_1$ when $-1<(g_0+I_0)<1/p_2$ such that the Chua circuit can operate at the origin $(0,0,0)$ or have a unstable operation according to the initial conditions.
The Nyquist diagram $G(j\omega)$ intercepts $-1/N(X)$ in unstable equilibrium points $\pm p_1$ and stable limit cycle $p_2$ for $1/p_2<(g_0+I_0)<1/p_3$, and the Chua circuit can have a unstable operation or oscillate in a stable limit cycle with frequency $\omega_2$ that may evolve to chaos as result of a possible interaction between interception points.
Since the unstable limit cycle $p_3$ is also an interception point for $1/p_3<(g_0+I_0)<0$, the Chua circuit can operate at the origin $(0,0,0)$, oscillate in a stable limit cycle with frequency $\omega_2$ that may evolve to chaos, or have a unstable operation. 
The Chua circuit operates at the origin $(0,0,0)$ for $0<(g_0+I_0)<+\infty$ because $G(j\omega)$ does not involve $-1/N(x)$. 

\begin{figure}
\centering
\small
\begin{tabular}{cc}
\includegraphics[height=3cm]{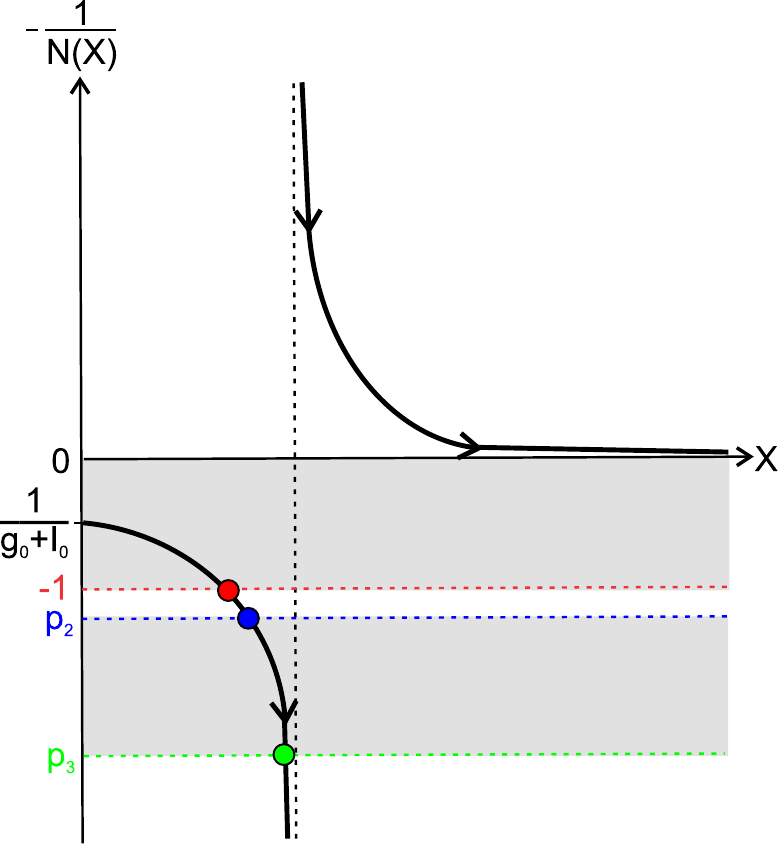} & \includegraphics[height=3cm]{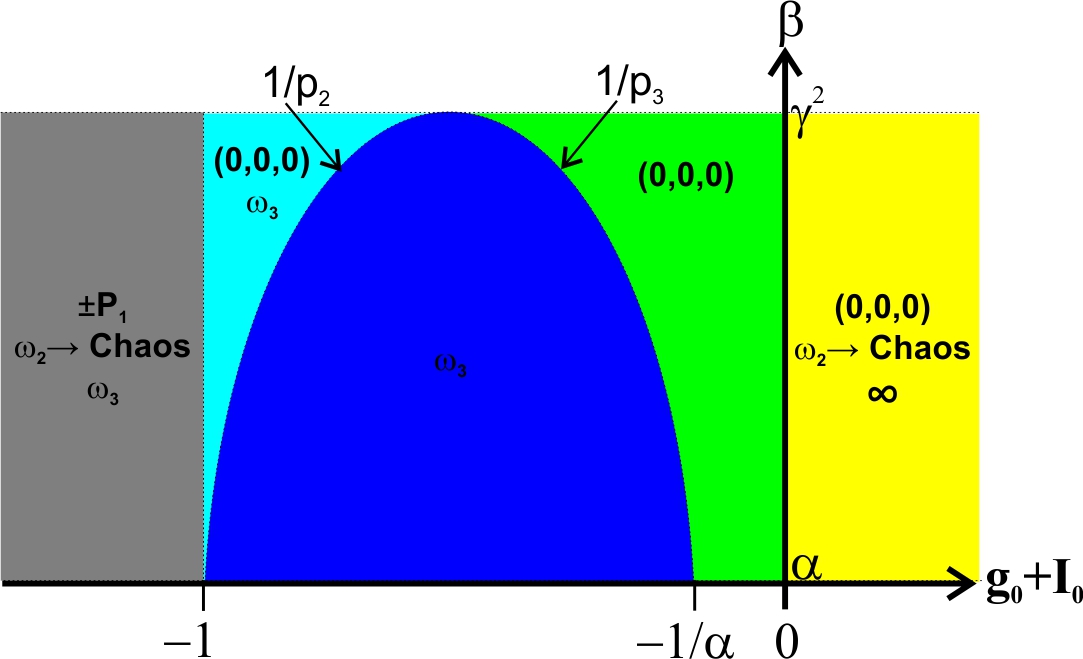}\\
(a)  & (b) 
\end{tabular}
\caption{Stability analysis for $\alpha<\beta<\gamma^2$ and $(g_0+I_0) I_0<0$ using describing functions. (a) The geometric locus $-1/N(X)$ superposed to the Nyquist diagram $G(j\omega)$ and the three possible interception points: $\pm p_1$ (red circle), $p_2$ (blue circle), and $p_3$ (green circle). 
(b) Sketch of the theoretical parameter space mapping $(g_0+I_0) \times \beta$ based on stability analysis, where $(0,0,0)=$ operation at origin $(0,0,0)$, $\pm P_1=$ operation at equilibrium points $\pm p_1$, $\omega_2\rightarrow Chaos=$ operation in a stable limit cycle with frequency $\omega_2$ that can evolve to chaotic behavior,  $\omega_3=$ operation in a stable limit cycle with frequency $\omega_3$, and $\infty=$ unstable circuit.}
\label{Fig:summary}
\end{figure}

Since the describing function $N(x)$ has real roots for $(g_0+I_0)I_0<0$, the geometric locus $-1/N(X)$ is a discontinuous real function that starts at $-1/(g_0+I_0)$ for $X=0$ and ends at $0$ for $X=\infty$ as shown Fig. \ref{Fig:summary}.(a).
The geometric locus  $-1/N(X)$ intercepts $G(j\omega)$ in stable equilibrium points $\pm p_1$, unstable limit cycle $p_2$, and stable limit cycle $p_3$ for $-\infty<(g_0+I_0)<-1$, such that Chua circuit can operate in the stable equilibrium points $\pm p_1$, oscillate in the stable limit cycle with frequency $\omega_3$, or present an oscillation with frequency $\omega_2$ that may evolve to chaos due to possible interactions between $\pm p_1$ and $p_2$.
The Nyquist diagram $G(j\omega)$ intercepts $-1/N(X)$ in unstable limit cycle $p_2$ and stable limit cycle $p_3$ for $-1<(g_0+I_0)<1/p_2$, such that the Chua circuit can operate at origin $(0,0,0)$ or oscillate in the stable limite cycle with frequency $\omega_3$.
The only interception point between $-1/N(x)$ and $G(j\omega)$ is $p_3$ when $1/p_2<(g_0+I_0)<1/p_3$, such that the Chua circuit can only oscillates in the stable limit cycle with frequency $\omega_3$.
The Chua circuit operates at origin $(0,0,0)$ for $1/p_3<(g_0+I_0)<0$ because $G(j\omega)$ does not involve $-1/N(X)$ in this parameter range.
The geometric locus $-1/N(X)$ intercepts $G(j\omega)$ in unstable equilibrium point $p_1$, stable limit cycle $p_2$, and unstable limit cycle $p_3$ for $0<(g_0+I_0)<+\infty$, such that the Chua circuit can operate at origin $(0,0,0)$, oscillate in a stable limit cycle with frequency $\omega_2$ that may evolve to a chaotic attractor, or have a unstable operation.
The results of the stability analysis of the Chua circuit for $\alpha<\beta<\gamma^2$ and $(g_0+I_0)I_0<0$ are summarized in the sketch of the parameter space mapping presented in Fig. \ref{Fig:summary}.(b).

\section{Numerical Analysis}\label{sec:num}
\begin{figure}
\centering
\small
\includegraphics[height=7cm]{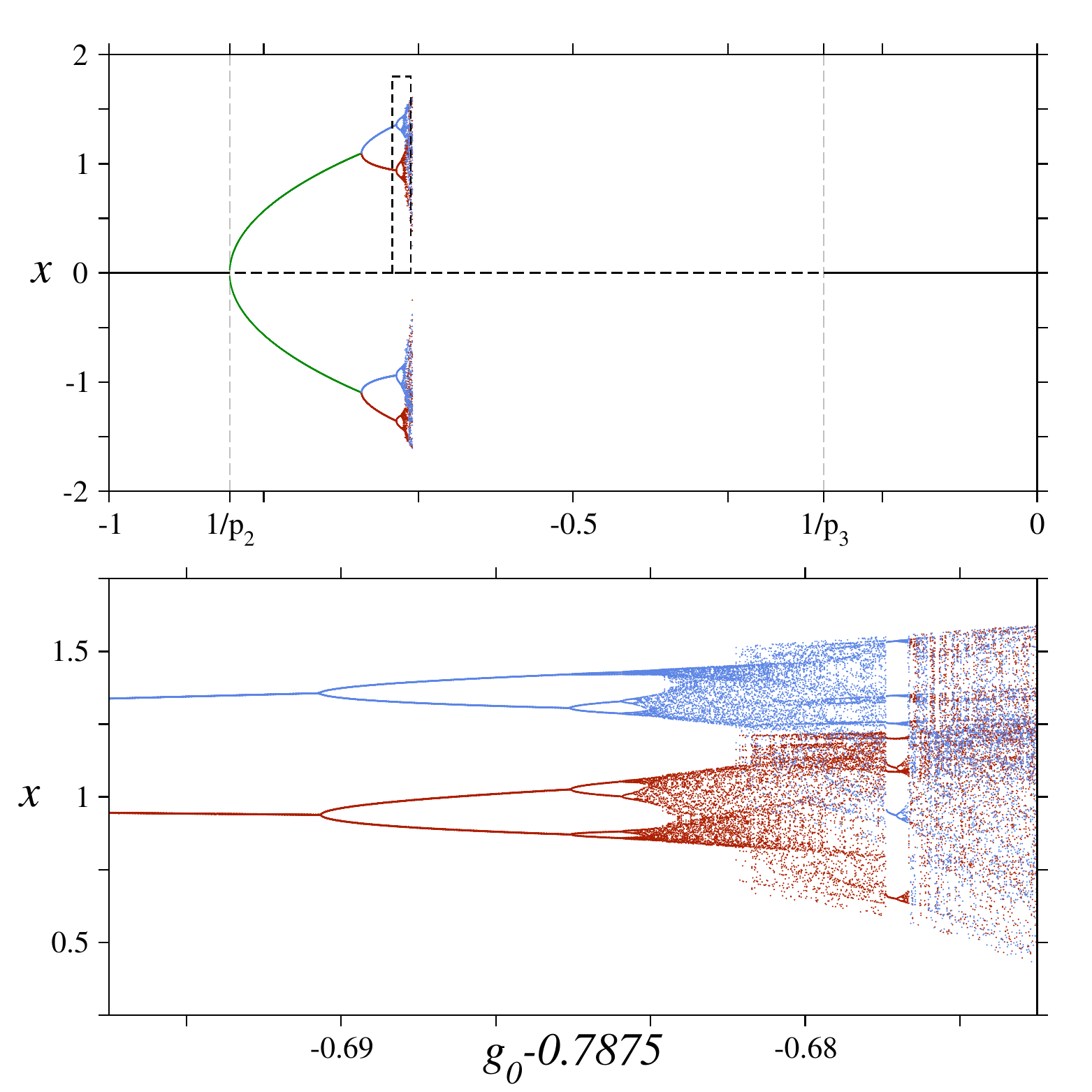}
\caption{Bidirectional bifurcation diagram for hyperbolic Chua circuit with $(g_0+I_0)I_0>0$ as $g_0$ varies: $\alpha = 10$, $\beta = 20$, $I_0=-0.7875$, and $-0.2125\leq g_0 \leq 0.7875$.
Dashed gray lines are the interception points $1/p_2 = -0.87$ and $1/p_3 = -0.23$. 
Black line indicates when the equilibrium at the origin is stable (continuous) and unstable (dashed). 
The green line represents the single limit cycle, while brown and blue colors represent the branches of coexisting symmetric attractors. 
The dashed square region is magnified in the bottom.}
\label{Fig:bifurcation}
\includegraphics[height=4cm]{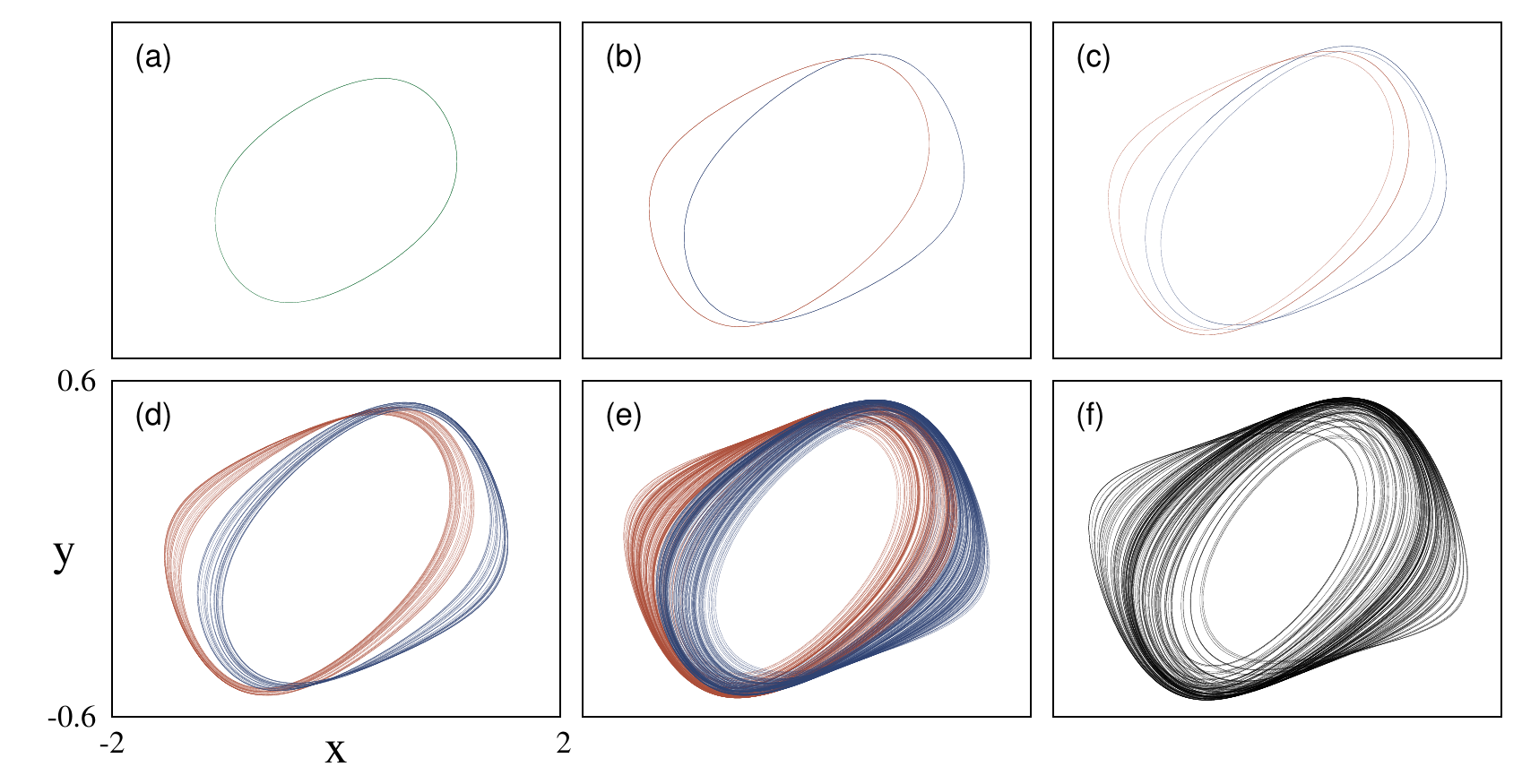}
\caption{Attractors evolution in hyperbolic Chua circuit with $(g_0+I_0)I_0>0$ as $g_0$ varies: $\alpha=10$, $\beta=20$ and $I_0=-0.7875$. (a) $g_0+I_0 =-0.75$, (b) $g_0+I_0=-0.7$, (c) $g_0+I_0=-0.687$, (d) $g_0+I_0=-0.682$, (e) $g_0+I_0=-0.68$, and (f) $g_0+I_0=-0.676$. 
All attractors are in the same scale shown in (d). 
The colors follow the bifurcation diagram in Fig. \ref{Fig:bifurcation}.}
\label{Fig:attractor}
\end{figure}

The features of the asymptotic behavior of the Chua circuit with nonlinearity based on antiparallel connected semiconductor diodes for $(g_0+I_0) I_0>0$ are presented in numerical bidirectional bifurcation diagram in Fig. \ref{Fig:bifurcation}, where are considered trajectories crossing the Poincar\'e section at $y=0$ in both directions (from negative $y$ to positive $y$ and vice versa) in order to observe symmetry between attractors.
Considering $\alpha = 10$, $\beta = 20$, and $I_0 = -0.7875$, the value of $g_0$ is varied from $-0.2125$ to $0.7875$ such that the bifurcation diagram spans the range $-1\leq (g_0+I_0) \leq 0$.
As $g_0$ increases, the equilibrium at origin (solid black line) is stable up to $(g_0+I_0) = 1/p_2$, when becomes unstable (dashed black line) due to an Andronov-Hopf bifurcation that gives rise to the limit cycle with frequency $\omega_2$ (two green branches) shown in Fig. \ref{Fig:attractor}.(a).
A pitchfork bifurcation of periodic oscillation splits this oscillation in two new limit cycles with frequency $\omega_2$ (blue and brown branches), which are shown in Fig. \ref{Fig:attractor}.(b).
The detail of the bifurcation diagram is presented in the bottom of Fig. \ref{Fig:bifurcation}, which shows the occurrence of a cascade of period doubling bifurcations in each branch starting from the double period attractors shown in Fig. \ref{Fig:attractor}.(c).
The dynamics becomes chaotic after $(g_0+I_0) = -0.6835$.
After a crisis, chaotic single scroll attractors presented in Fig. \ref{Fig:attractor}.(d) suddenly enlarge their sizes as shown in Fig. \ref{Fig:attractor}.(e), and collide each other to form only one chaotic single scroll attractor, which is presented in Fig. \ref{Fig:attractor}.(f).
Cascades of period doubling bifurcations are also replicated in windows of periodic oscillations that arise inside the range of the chaotic behavior.
The system loses stability as $g_0$ increases and only restores to equilibrium at the origin for $g_0+I_0\geq 1/p_3$.

\begin{figure}
\centering
\small
\includegraphics[height=7cm]{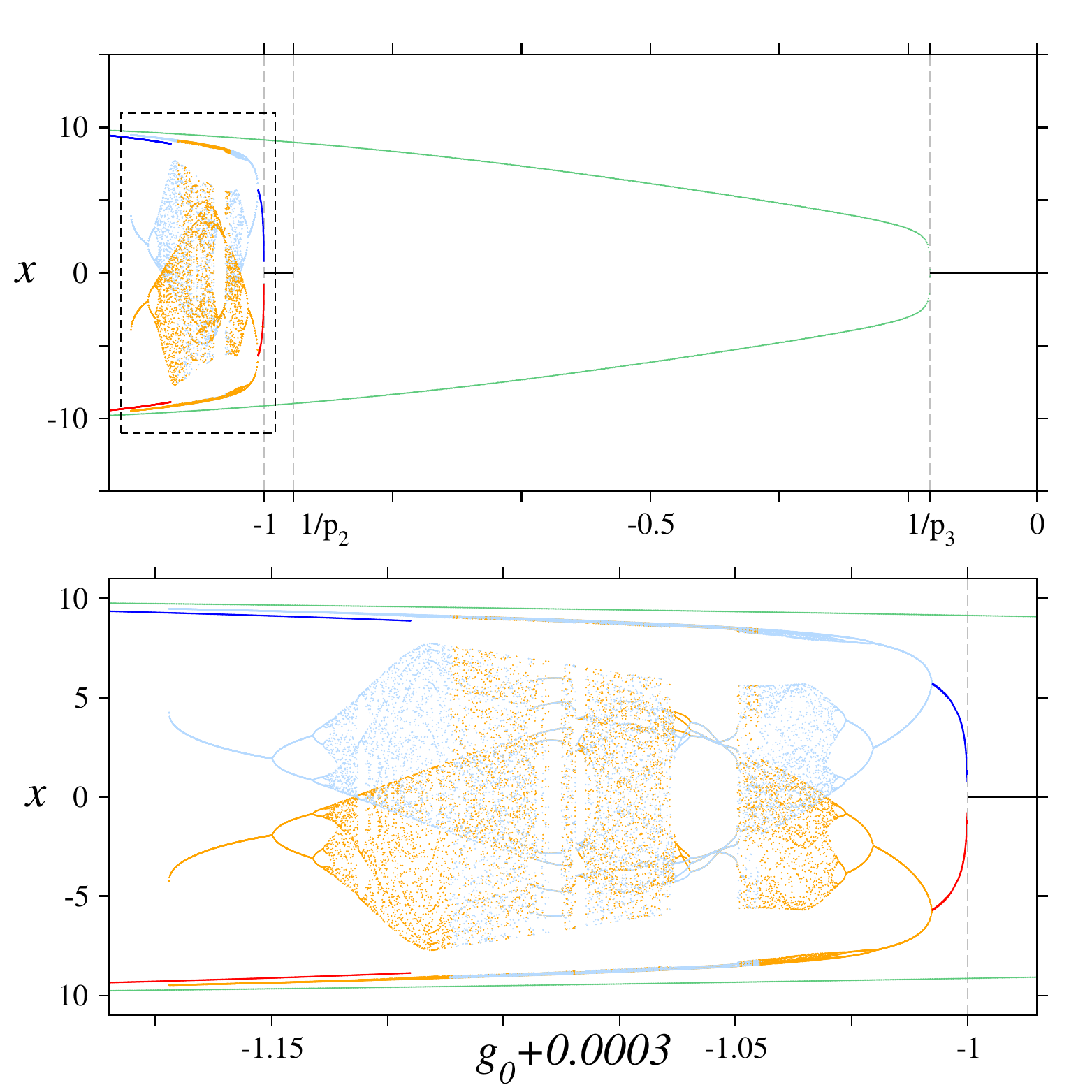}
\caption{Bidirectional bifurcation diagram in hyperbolic Chua circuit with $(g_0+I_0)I_0<0$ as $g_0$ varies: $\alpha = 10$, $\beta = 13.3$, $I_0=0.0003$, and $-1.2003\leq g_0 \leq -0.0003$. 
Dashed gray lines are the interception points $1/p_1=-1$, $1/p_2 = -0.96$, and $1/p_3 = -0.14$.
Black line indicates when the equilibrium at the origin is stable (continuous).
The green lines represents the limit cycle with frequency $\omega_3$.
Red, orange, blue, and light blue branches represent coexisting symmetric attractors. 
The dashed square region is magnified in the bottom.}
\label{Fig:bifurcationb}
\includegraphics[height=6cm]{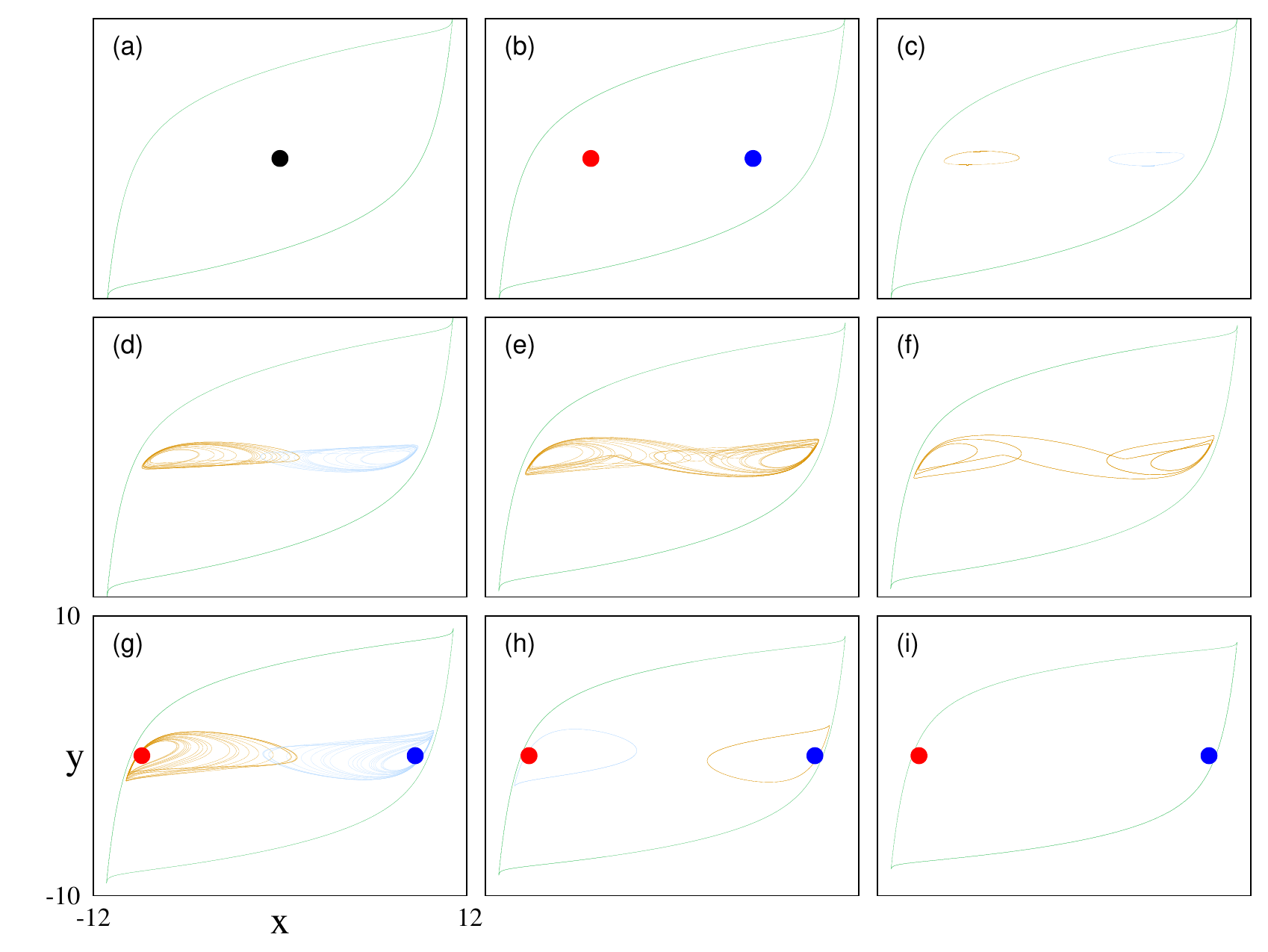}
\caption{Attractors evolution in the Chua circuit with hyperbolic characteristic and $(g_0+I_0)I_0<0$ as $g_0$ varies: $\alpha = 10$, $\beta = 13.3$ and $I_0= 0.0003$.
(a) $g_0+I_0 =-0.99$, (b) $g_0+I_0 =-1.005$, (c) $g_0+I_0 =-1.017$, (d) $g_0+I_0 =-1.04$, (e) $g_0+I_0 =-1.07$, (f) $g_0+I_0 =-1.09$, (g) $g_0+I_0 =-1.12$, (h) $g_0+I_0 =-1.16$, and (i) $g_0+I_0 =-1.18$. All attractors are in the same scale shown in (g). The colors follow the bifurcation diagram in Fig. \ref{Fig:bifurcationb}.}
\label{Fig:attractorb}
\end{figure}

A numerical bidirectional bifurcation diagram for Chua circuit with hyperbolic characteristic and $(g_0+I_0)I_0<0$ for $\alpha = 10$, $\beta = 13.3$, $I_0 = 0.0003$, and  $-1.2003<g_0<-0.0003$ is presented in Fig. \ref{Fig:bifurcationb}.
The equilibrium at origin (black line) is initially stable.
As $g_0$ decreases, an Andronov–Hopf bifurcation occurs in $(g_0+I_0)=1/p_3$ and the Chua circuit begins periodically oscillate in a limit cycle with frequency $\omega_3$ (green branches).
This limit cycle becomes a hidden oscillation from $(g_0+I_0)=1/p_2$, when it starts to coexist with stable equilibrium points, periodic oscillations and chaotic dynamics as shown in Figs. \ref{Fig:attractorb}.(a) to (i).
The stability of the origin is rescued from $(g_0+I_0)=1/p_2$ to $(g_0+I_0)=-1$, and gives place to two stable symmetrical equilibrium points $\pm P_1$ (blue and red lines) in a typical scenery of super-critical pitchfork bifurcation as shows the detail at the bottom of Fig. \ref{Fig:bifurcationb} and Figs. \ref{Fig:attractorb}.(a) and (b).
These symmetrical equilibrium points $\pm P_1$ become simultaneously unstable due to Andronov–Hopf bifurcations, and self-excited periodic oscillations with frequency $\omega_2$ (light blue and orange branches) emerge in this process.
These self-excited periodic oscillations in Fig. \ref{Fig:attractorb}.(c) evolve to twin R\"ossler type attractors in Fig. \ref{Fig:attractorb}.(d) via cascades of period doubling bifurcations in a route to chaos known as Feigenbaum scenery, which collide each other in order to form the well-known double scroll attractor in Fig. \ref{Fig:attractorb}.(e).
Windows with periodic attractors such as that shown in Fig. \ref{Fig:attractorb}.(f) arise inside the range of the chaotic behavior, where are also replicated routes to chaos with cascades of period doubling bifurcations.
In an inverse process of cascade of period doubling bifurcations, the double scroll attractor splits back to two twin R\"ossler type attractors, which become twin periodic oscillations $\omega_2$ as shown in Figs. \ref{Fig:attractorb}.(g) and (h).
These dynamics become hidden oscillations after the equilibrim points $\pm P_1$ rescue their stability as shown in Fig. \ref{Fig:attractorb}.(g), and lose stability as $g_0$ decreases.
The Chua circuit operates in the symmetrical equilibrium points $\pm P_1$ or oscillates in the hidden limit cycle with frequency $\omega_3$ as shown in Fig. \ref{Fig:attractorb}.(i) for more negative values of $(g_0+I_0)$.

\section{Implementations of Chua Circuit with Exponential Hyperbolic Characteristics} \label{sec:electronics}
\begin{figure}
\centering
\small
\includegraphics[height=2.5cm]{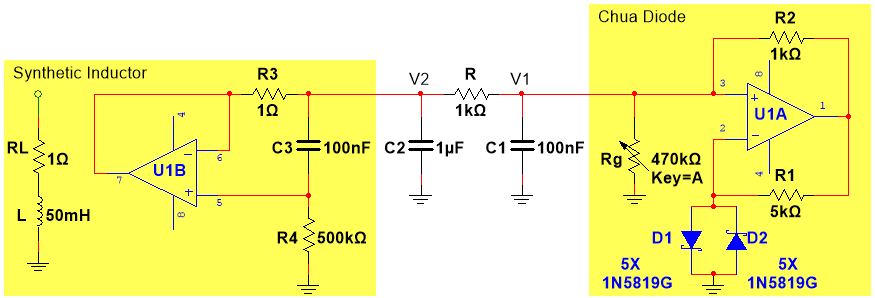}
\caption{Inductorless  exponential hyperbolic  Chua circuit with $(g_0+I_0)I_0>0$: $\alpha=10$, $\beta=20$, $g_0>1/470k$, and $I_0\approx-0.7875$.}
\label{Fig:electronic_1}
\centering
\small
\begin{tabular}{ccc}
\includegraphics[width=.125\linewidth]{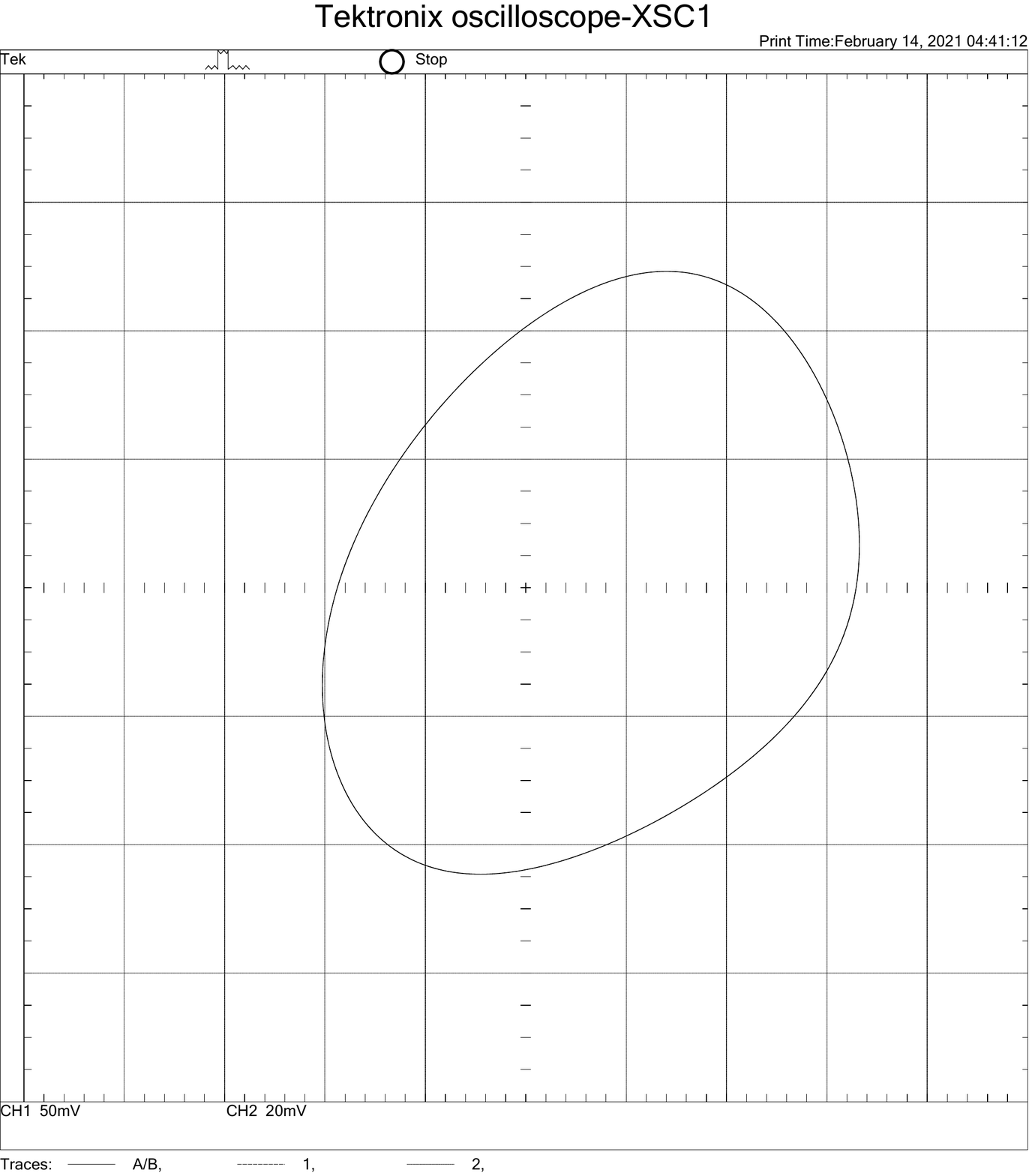}&\includegraphics[width=.125\linewidth]{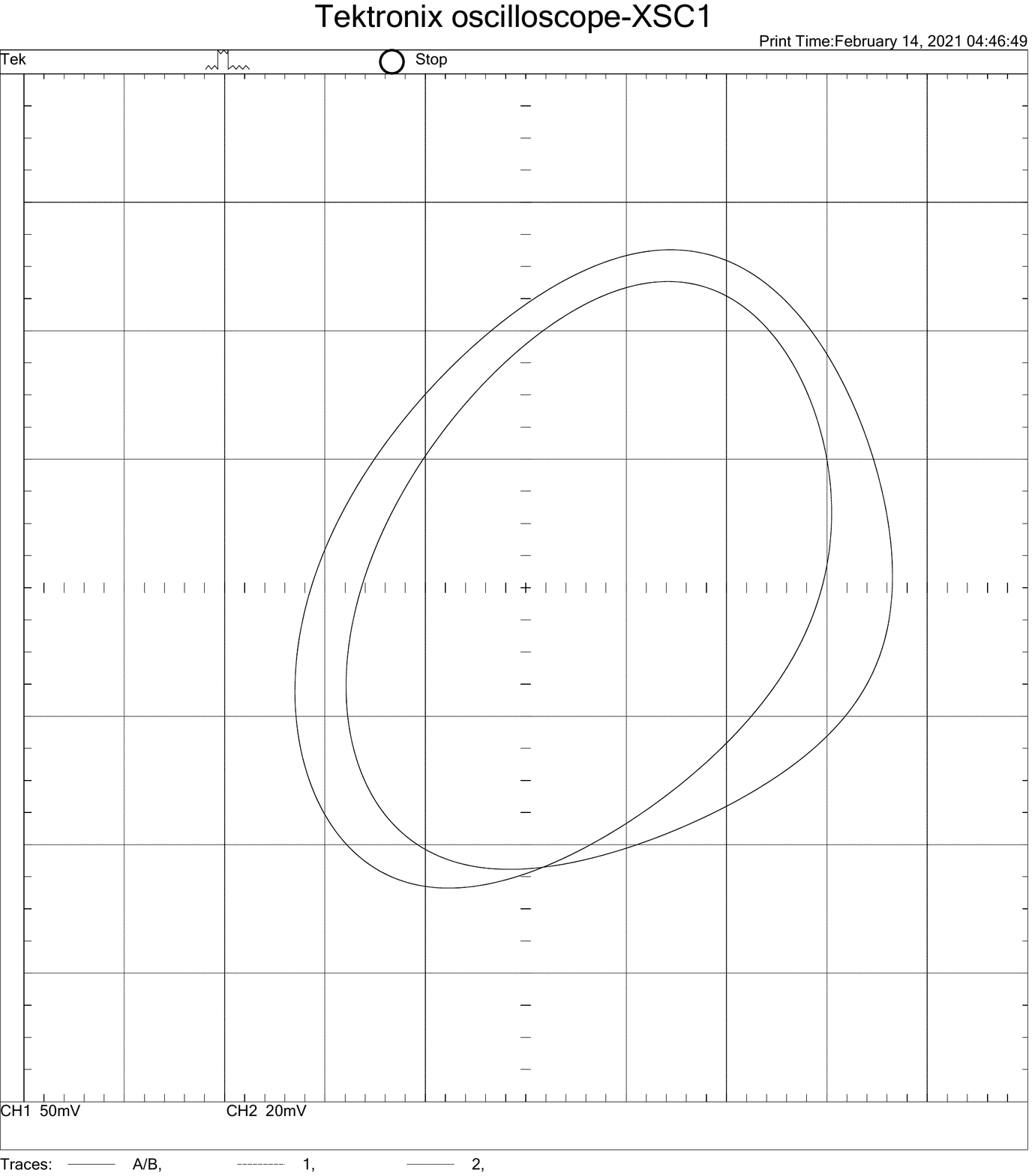}& \includegraphics[width=.125\linewidth]{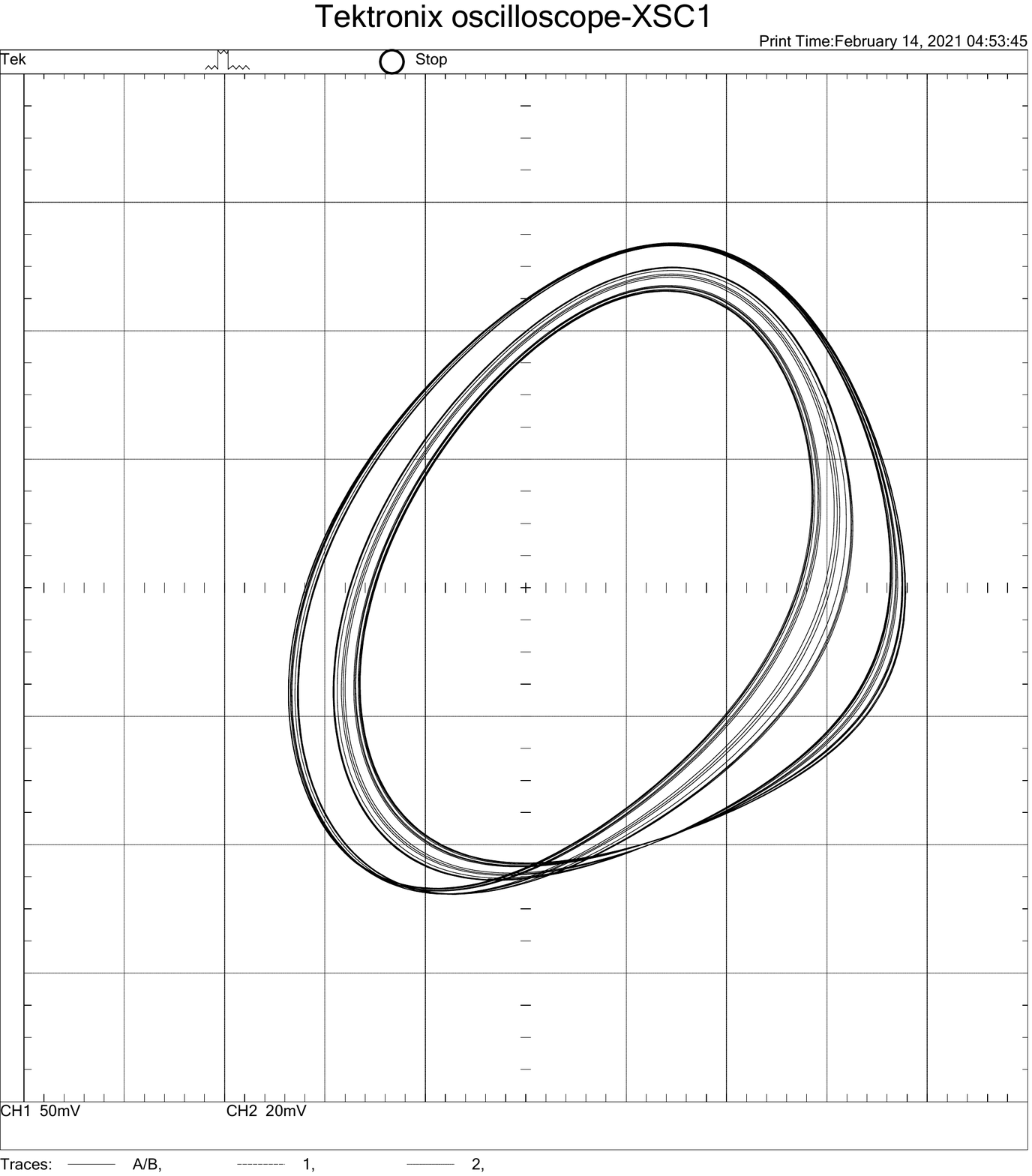}\\
(a) &(b)  & (c)\\
\includegraphics[width=.125\linewidth]{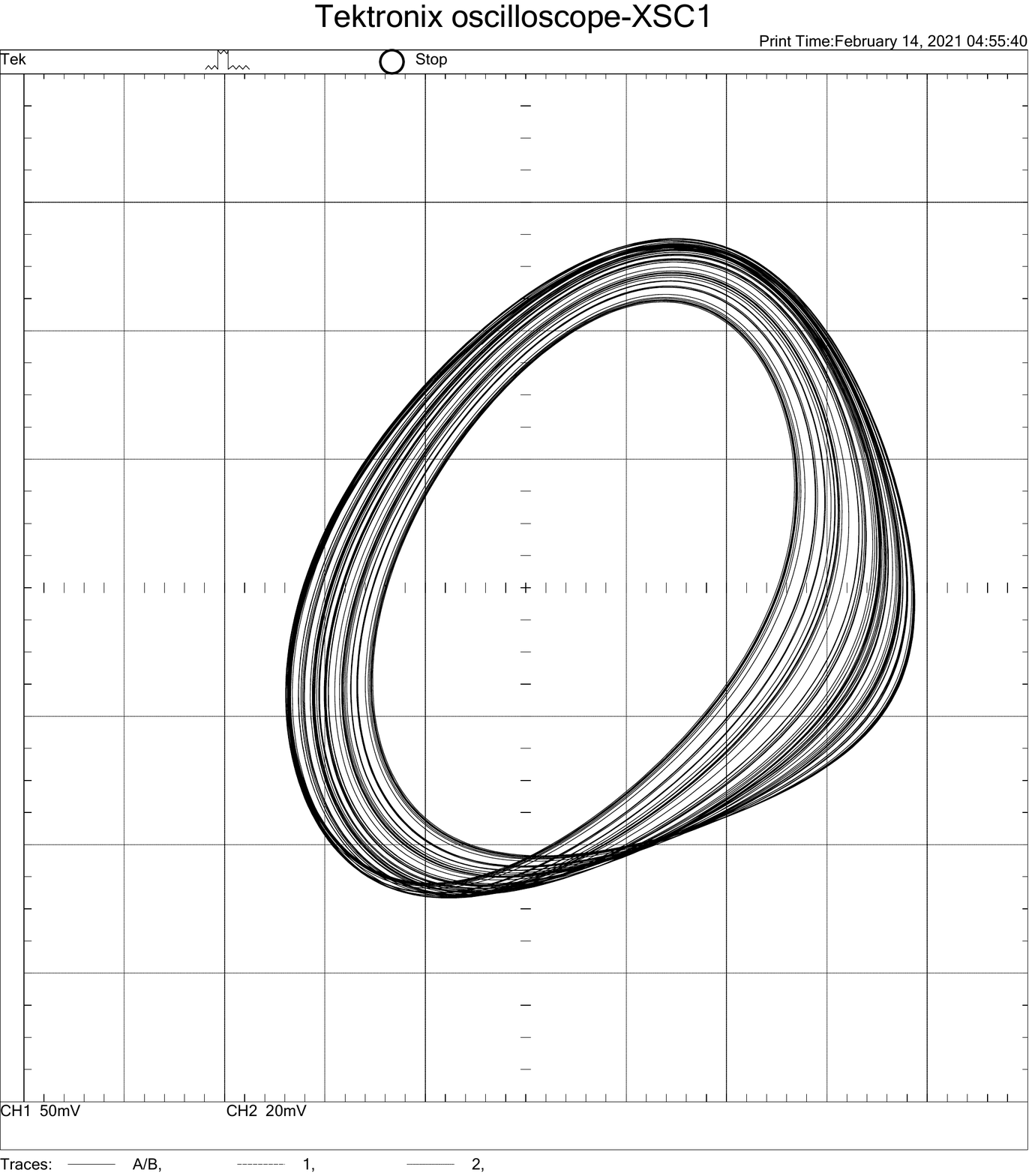}&\includegraphics[width=.125\linewidth]{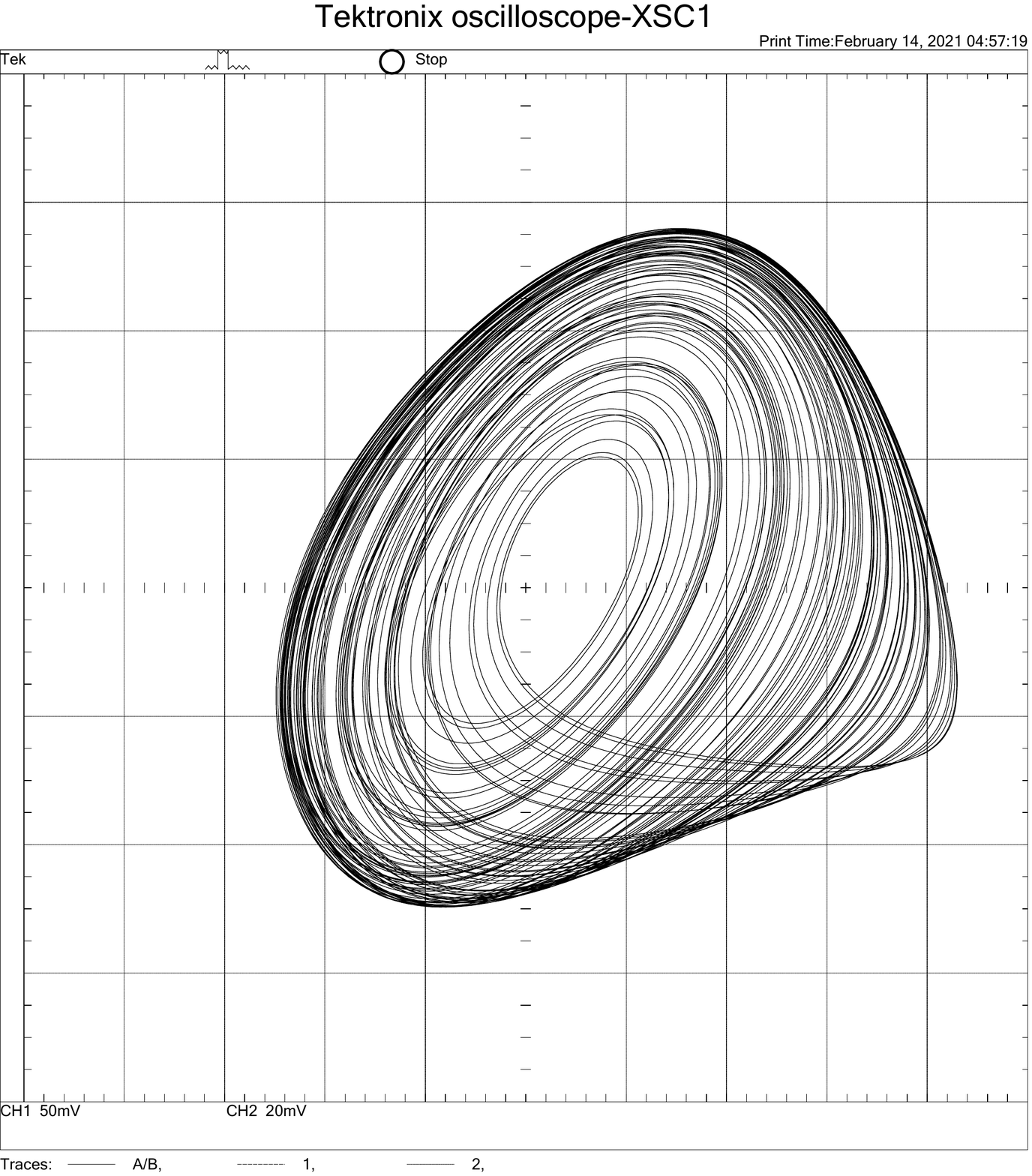}& \includegraphics[width=.125\linewidth]{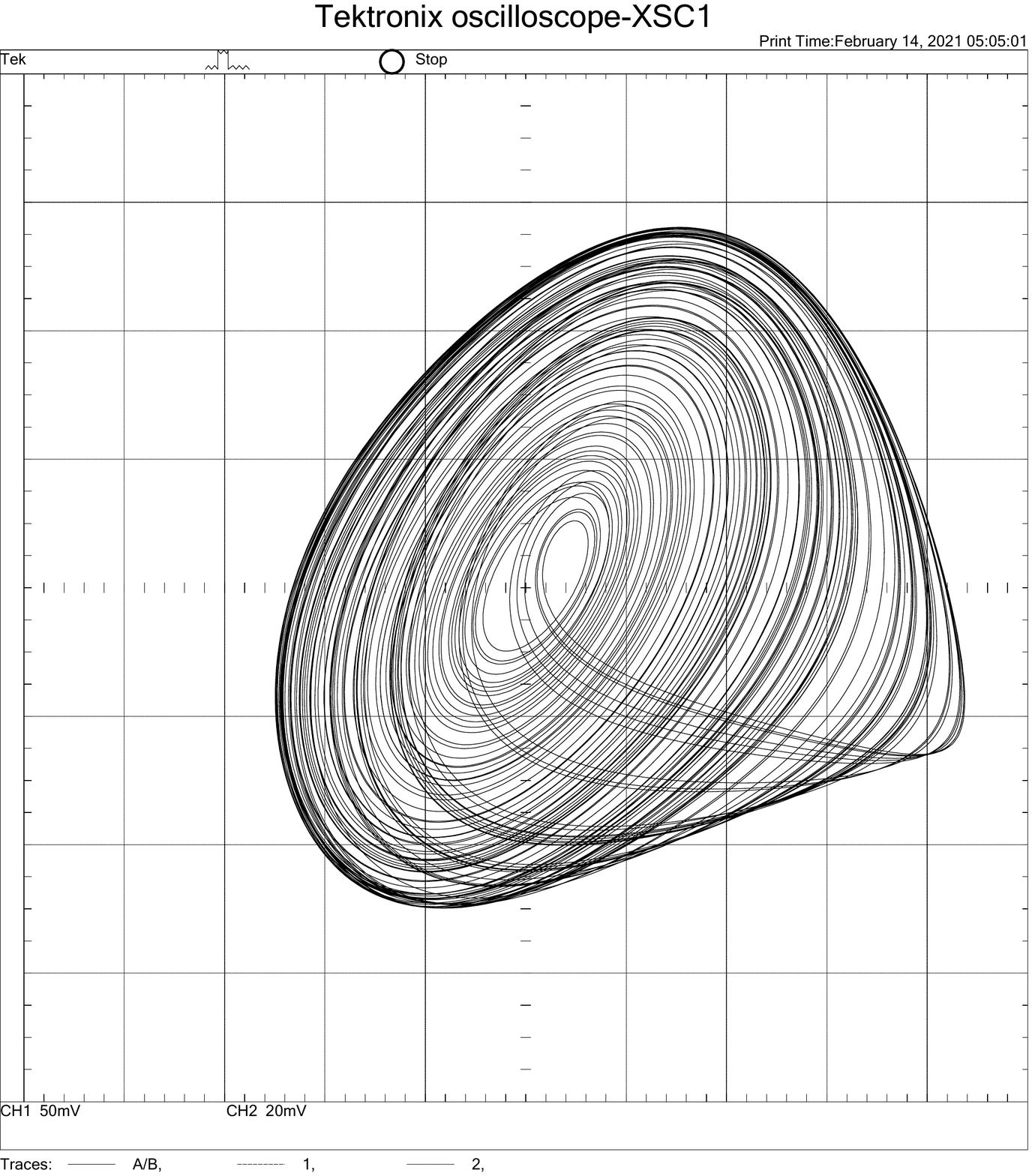}\\
(d) &(e)  & (f)
\end{tabular}
\caption{Evolution of $v_2-v_1$ attractors obtained in simulations of the inductorless Chua circuit with active antiparallel connection of semiconductor diodes: (a) $g_0=0.0021$, (b) $g_0=0.0100$, (c) $g_0=0.0143$, (d) $g_0=0.0167$, (e) $g_0=0.0200$, and (f) $g_0=0.0208$. Amplitude scale for (a) to (i): CH1=50mV/div ($v_1$) and CH2=20mV/div ($v_2$). }
\label{Fig:exp_attractor_1}
\end{figure}

The implementation of an inductorless Chua circuit with $(g_0+I_0)I_0>0$ using a dual low noise JFET-input op-amps TL072 is presented in Fig. \ref{Fig:electronic_1}. 
The active device with hyperbolic characteristic is a negative impedance converters with $i-v$ characteristic given by

\begin{equation}
i=-2\frac{R_2}{R_1}li_s\sinh\left(\frac{v}{m\eta v_t}\right),
\end{equation}
which is provided by the antiparallel connection of two sets of five serie-connected Schottky diodes 1N5819.
The real parameters of commercially available discrete semiconductor diodes can be difficult to evaluate since they depend on circuit conditions and vary significantly among same type components, such that the reverse current and emission coefficient of the Schottky diode 1N5819 are respectively considered as 
$i_s=11.928\mu A$ and $\eta=1.165$ for computational simulation purposes.
An implementation of a Chua diode with exponential hyperbolic characteristics can require semiconductor diodes with high reverse current at low voltages, which may be bypassed using parallel connected semiconductor diodes in order to obtain adequate levels of $i_s$.
The resistances $R_1=5k\Omega$ and $R_2=1k\Omega$ are selected such that $I_0\approx -0.7875$ for $R=1k\Omega$. 
A rheostat $R_g=470k\Omega$ varies the passive parallel conductance $g_0$, which assures the condition $(g_0+I_0)I_0>0$.
The capacitances are selected as $C_1=100nF$ and $C_2=1\mu F$ in order to obtain $\alpha=10$ and $\tau=1m$s.
An inductance $L=50mH$ with negligible series resistance is synthesized in order to obtain $\beta=20$ using a single op-amp synthetic inductor circuit whose impedance is approximately given by \citep{horowitz-1989,muthuswamy-2009}

\begin{equation}
Z \approx R_3+j\omega R_3R_4C_3=R_L+j\omega L,
\end{equation}
where $R_3=1\Omega$, $R_4=500k\Omega$, and $C_3=100nF$.
This inductorless exponential hyperbolic Chua circuit with $(g_0+I_0)I_0>0$ is simulated using the software NI MultiSim\texttrademark 13 and the evolution of $v_2-v_1$ attractors as $R_g$ varies is presented in Fig. \ref{Fig:exp_attractor_1}, where is observed that an initial stable limit cycle suffers a cascade of period doubling bifurcation as $g_0$ increases until becomes a chaotic single scroll attractor.
An experimental implementation of this version of Chua circuit is infeasible in this work due to the low amplitudes of the generated signals, which are difficult to observe in conventional CRT oscilloscopes and can be easily corrupted by noises. 
This problem may be bypassed using a large quantity of series-connected semiconductor diodes to obtain adequate voltage amplitudes and assure an acceptable signal-to-noise ratio.

\begin{figure}
\centering
\small
\includegraphics[height=2.5cm]{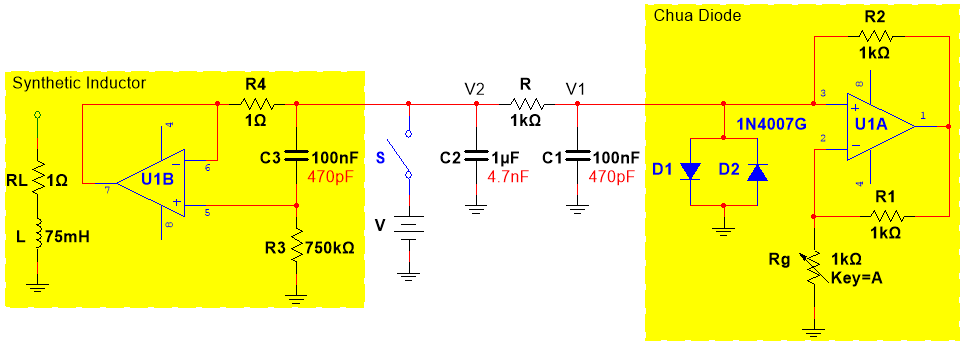}
\caption{Inductorless  exponential hyperbolic Chua circuit with $(g_0+I_0)I_0<0$, $\alpha=10$, $\beta\approx 13.3$, $g_0<-1.01$, and $I_0\approx 0.0003$.}
\label{Fig:electronic_2}
\centering
\small
\begin{tabular}{ccc}
\includegraphics[width=.2\linewidth]{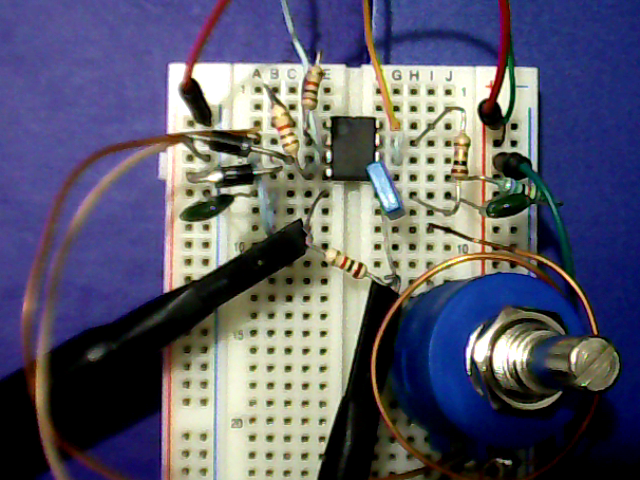} & \includegraphics[width=.2\linewidth]{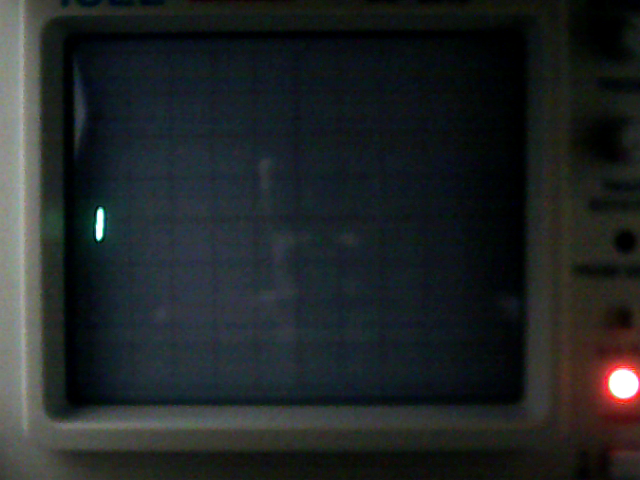}&\includegraphics[width=.2\linewidth]{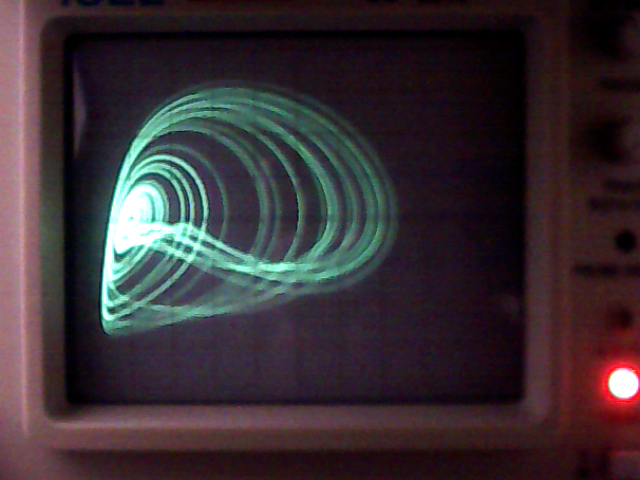}\\
(a) & (b) & (c)\\
\includegraphics[width=.2\linewidth]{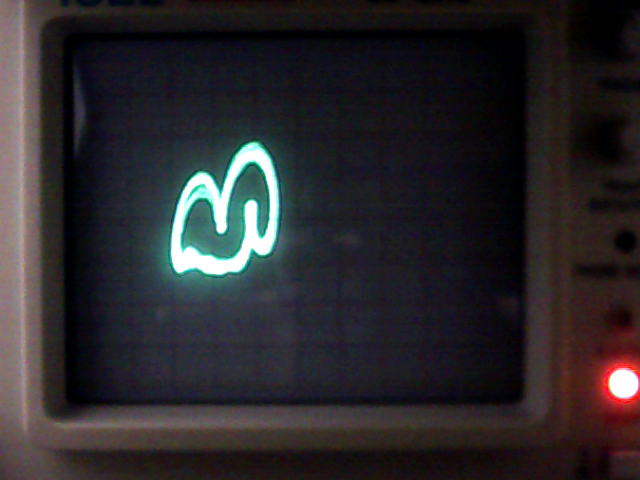} &\includegraphics[width=.2\linewidth]{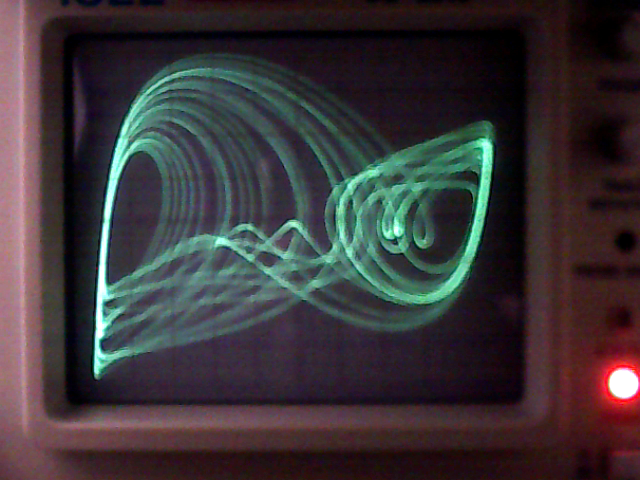} & \includegraphics[width=.2\linewidth]{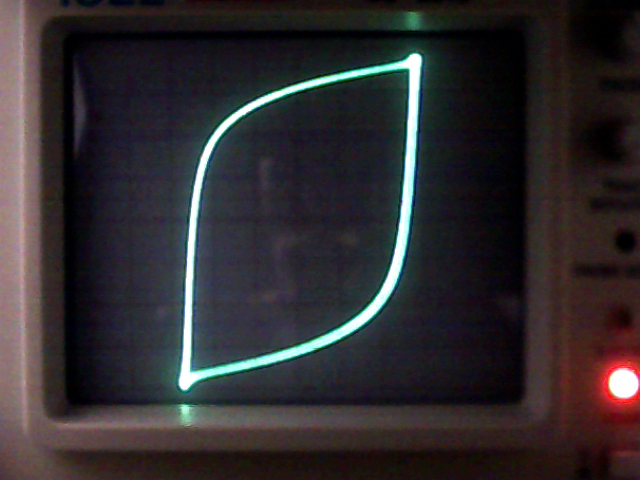}\\
 (d)& (e) & (f)\\
\end{tabular}
\caption{Experimental $v_2-v_1$ attractors generated by the experimental implementation of the inductorless Chua circuit with $(g_0+I_0)I_0<0$ and slower dynamics ($\alpha=10$, $\beta\approx 13.3$, $\tau=1m$s, $g_0<-1.01$, and $I_0\approx 0.0003$) observed in a 20MHz analog oscilloscope on X-Y mode. (a) Experimental prototype, (b) Equilibrium point, (c) R\"ossller-type attractor, (d) Periodic attractor, (e) Double scroll attractor, and (f) Hidden oscillation.}
\label{Fig:fotos2}
\centering
\small
\begin{tabular}{ccc}
\includegraphics[width=.2\linewidth]{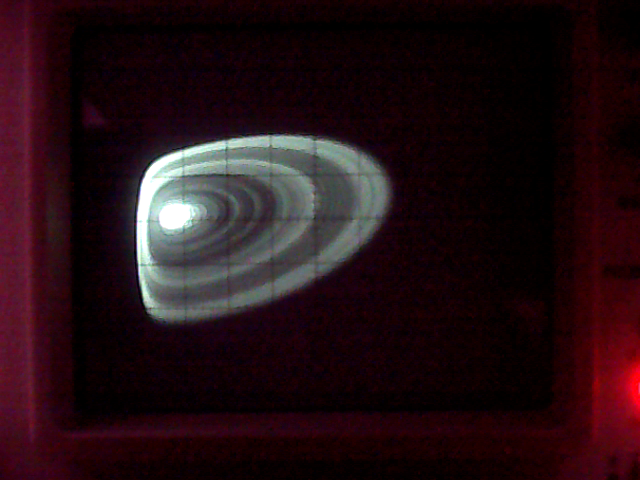}&\includegraphics[width=.2\linewidth]{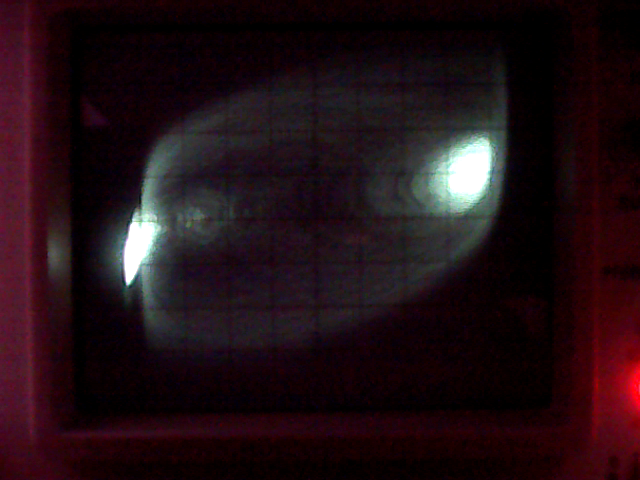}&\includegraphics[width=.2\linewidth]{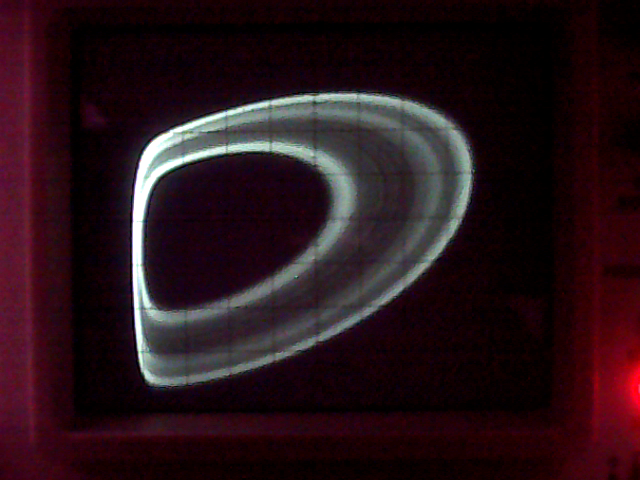}\\
(a)  & (b) & (c)  \\
\includegraphics[width=.2\linewidth]{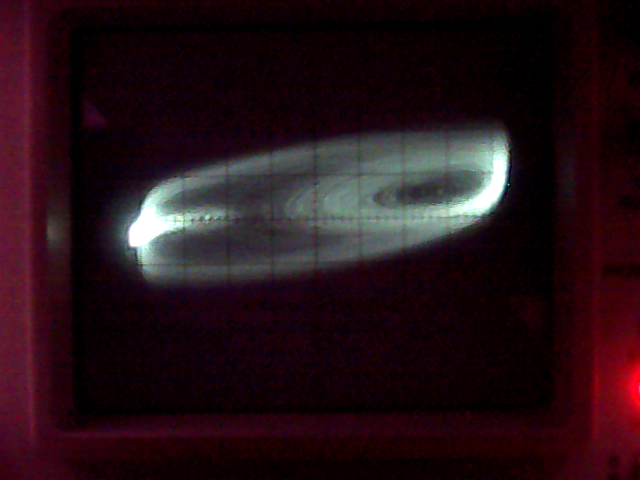} &\includegraphics[width=.2\linewidth]{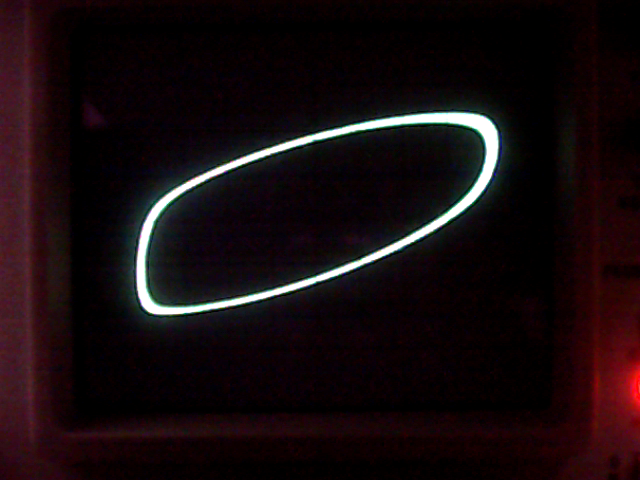}&\includegraphics[width=.2\linewidth]{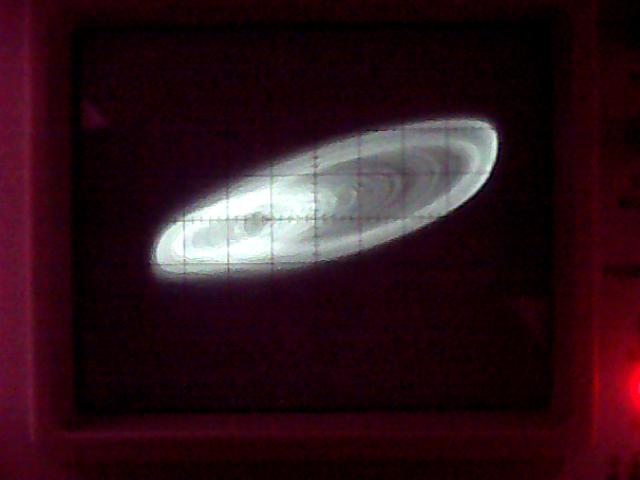}\\
(d)&(e) &(f) 
\end{tabular}
\caption{Experimental $v_2-v_1$ attractors generated by the experimental implementation of the inductorless Chua circuit with $(g_0+I_0)I_0<0$ and fastest dynamics ($\alpha=10$, $\beta\approx 13.3$, $\tau=4.7\mu$s, $g_0<-1.01$, and $I_0\approx 0.0003$) observed in a 20MHz analog oscilloscope on X-Y mode.}
\label{Fig:photos3}
\end{figure}

The Chua circuit with $(g_0+I_0)I_0<0$ using a dual low noise JFET-input op-amps TL072 is presented in Fig. \ref{Fig:electronic_2}.
The active variable parallel conductance is a negative impedance converter, whose $i-v$ characteristic is given by

\begin{equation}
i=-\frac{R_2}{R_1R_g}v.
\end{equation}
where $R_1=R_2=1k\Omega$ and $R_g=1k\Omega$ is a rheostat. 
The hyperbolic nonlinearity is obtained from a passive antiparallel connection of a pair of standard low speed rectifier silicon diodes 1N4007.
The reverse current and emission coefficient of standard rectifier semiconductor diodes 1N4007 are respectively considered as $i_s = 7.061nA$ and $\eta= 1.808$ for design purposes.
The linear resistor is $R=1k\Omega$, such that $g_0<-1$, $I_0\approx 3\times 10^{-4}$, and $(g_0+I_0)I_0<0$.
The capacitances are selected as $C_1=100nF$ and $C_2=1\mu F$ in order to obtain $\alpha=10$ and $\tau=1m$s.
The single op-amp synthetic inductor circuit with $R_3=1\Omega$, $R_4=750k\Omega$, and $C_3=100nF$ synthesizes an inductance $L=75mH$ with negligible series resistance in order to obtain $\beta=13.3$.
The estimated frequencies for the oscillations $\omega_2$ and $\omega_3$ are respectively $f_2=307$Hz and $f_3=550$Hz, which are compatible with the specifications of the standard low speed rectifier silicon diodes 1N4007.
An on-off switch $S$ allows set alternative initial conditions to $C_2$ in order to observe hidden oscillations.
The experimental implementation in a base for prototyping of electronic circuits is shown in Fig. \ref{Fig:fotos2}.(a), whose experimentally generated $v_2-v_1$ attractors in Figs. \ref{Fig:fotos2}.(b) to (f) reproduce the main dynamics of the Chua circuit.

 {The exponential hyperbolic current-voltage characteristic is the predominant non-ideality of semiconductor diodes in low voltage and low frequency applications. 
However, real semiconductor diodes have other nonlinearities that are often neglected in designs and simulations, but they can become representative in certain operational circumstances. 
Since these unmodelled non-idealities are always present in the behavior of real semiconductor diodes and can influence the operation of the electronic circuit, it is interesting verify how they affect the operation of the experimental Chua circuit with hyperbolic nonlinearity. 
Since the effect of some unmodelled non-idealities of real semiconductor diodes are most significant in operations at high frequencies, the capacitors of the proposed Chua circuit are resized in order to obtain an experimental implementation using low speed diodes with the same dimensionless parameters and fastest oscillations.
The dynamics of this Chua oscillator with $(g_0+I_0)I_0<0$ is increased around two hundred times without changing the dimensionless parameters $\alpha=10$ and $\beta\approx 13.3$ with $C_1=C_3=470pF$ and $C_2=4.7nF$.
These capacitance changings reduces the time constant of the circuit to $\tau=4.7\mu$s such that the estimated frequencies for $\omega_2$ and $\omega_3$ rise respectively to $f_2=65$kHz and $f_3=117$kHz.
This experimental implementation with faster dynamics may present higher sensitivity to external energy fields since hot spots of physical prototype can operate as antennas. 
Some $v_2-v_1$ attractors experimentally generated by this Chua oscillator prototype with $(g_0+I_0)I_0<0$ and fast dynamics are presented in Fig. \ref{Fig:photos3}.
These experimental results show that the dynamic behavior of experimental high speed implementation is different from that is expected for a conventional Chua circuit, which demonstrates a significant influence of the unmodelled non-idealities of semiconductor diodes in the operation of the proposed version of the Chua circuit.
This unconventional dynamic behavior of the fast Chua circuit is probably due to the electric charges stored in parasitic voltage dependent capacitances of diodes 1N4007, which cause delays and compromise the operation in high speed applications.
This problem could be solved substituting the low speed rectifier diodes 1N4007 by appropriate faster recovery time diodes in this experimental implementation.
In other hand, this problem can be converted into an opportunity for studies and researches of new possible dynamics in the Chua circuit exploring the unmodelled non-idealities of semiconductor devices.}

\section{Conclusions} \label{sec:conclusion}
 {This paper characterizes the dynamics of the Chua circuit with nonlinearity based on the inherent exponential hyperbolic characteristics of real semiconductor devices.
The dynamics of this nonlinear system is analytically predicted using the method of the describing functions and mapped in parameter space to create a base for studies, analyses, and designs. 
This theoretical analysis presents good agreement with numerical investigations and experimental results obtained with low speed circuit Chua circuit. 
However, the dynamic behavior of the experimental high speed circuit is different from that is expected for a conventional Chua circuit due to the effect of unmodelled non-idealities of the real semiconductor devices. 
This problem can be converted into an opportunity for studies and researches of new possible dynamics in the Chua circuit exploring the unmodelled non-idealities of semiconductor devices.}

\section*{Acknowledgement}
The author gratefully acknowledges
Coordination for the Improvement of Higher Level - or Education - Personnel (CAPES),
National Counsel of Technological and Scientific Development (CNPq), 
State of Minas Gerais Research Foundation (FAPEMIG), and State of S\~ao Paulo Research Foudation (FAPESP, Proc. 2015/50122-0).

\bibliographystyle{apalike}

\end{document}